**Classification:** Biological Sciences: Biophysics

# Universality and diversity of folding mechanics for three-helix bundle proteins

**Jae Shick Yang\*, Stefan Wallin\*, and Eugene I. Shakhnovich[†]**

Department of Chemistry and Chemical Biology, Harvard University, 12 Oxford Street, Cambridge, MA 02138;

\*These authors contributed equallly to this work.

[†]Correspondence should be directed to Prof. Eugene I. Shakhnovich, Department of Chemistry and Chemical Biology, Harvard University, 12 Oxford Street, Cambridge, MA 02138, phone: 617-495-4130, fax: 617-384-9228, e-mail:eugene@belok.harvard.edu

**Abbreviations:** MC, Monte Carlo; RMSD, rms deviation; DRMS, distance rms; MFPT, mean first passage time; MLET: mean last exit time, TSE: transition state ensemble.

# Abstract

In this study we evaluate, at full atomic detail, the folding processes of two small helical proteins, the B domain of protein A and the Villin headpiece. Folding kinetics are studied by performing a large number of *ab initio* Monte Carlo folding simulations using a single transferable all-atom potential. Using these trajectories, we examine the relaxation behavior, secondary structure formation, and transition-state ensembles (TSEs) of the two proteins and compare our results with experimental data and previous computational studies. To obtain a detailed structural information on the folding dynamics viewed as an ensemble process, we perform a clustering analysis procedure based on graph theory. Moreover, rigorous $p_{fold}$ analysis is used to obtain representative samples of the TSEs and a good *quantitative* agreement between experimental and simulated $\Phi$-values is obtained for protein A. $\Phi$-values for Villin are also obtained and left as predictions to be tested by future experiments. Our analysis shows that two-helix hairpin is a common partially stable structural motif that gets formed prior to entering the TSE in the studied proteins. These results together with our earlier study of Engrailed Homeodomain and recent experimental studies provide a comprehensive, atomic-level picture of folding mechanics of three-helix bundle proteins.

An eventual solution to the protein folding problem will involve a close calibration of theoretical methods to experimental data (1-4). In the endeavor of obtaining a *quantitative* agreement between theory and experiments, two small α-helical proteins have played a central role, namely the B domain of protein A from S*taphylococcus aurues* and the Villin headpiece subdomain from chicken. While these proteins belong to different SCOP fold classes (5), both have simple three-helix-bundle native topologies and fold autonomously on the μs time scale (6, 7). This makes them ideal test cases for protein simulations and numerous simulation studies, ranging from simple $C_\alpha$ Go-type to all-atom models with explicit water, have been undertaken for both protein A (8-21) and Villin (16, 17, 22-29).

Important advances have been made towards agreements with experiments for both proteins but several key issues remain unresolved (6, 30, 31). The need for additional studies is also emphasized by recent experiments. Fersht *et al.* (31, 32) performed a comprehensive mutational analysis on protein A by obtaining Φ-values at > 30 amino acid positions, providing a new important benchmark for simulation studies. The obtained Φ-values suggest that the transition-state ensemble (TSE) is characterized mainly by a well-formed H2 (we denote the three individual helices from N- to C-terminal by H1, H2, and H3, following previous convention.) stabilized by hydrophobic interactions with H1. Many simulation studies (8, 9, 11, 13, 16-18), although not all (10, 14), have emphasized H3 rather as the most stable helix and the first to form during folding, in line with early circular dichroism measurements on individual fragments of protein A (33). Recent experimental studies of Villin (34, 35) have focused mainly on achieving fast folding mutants, although new biophysical characterization of wild-type Villin was also obtained. Interestingly, the results indicate that these mutants are approaching the "speed-limit" for folding. Nonetheless, a limited free-energy barrier for folding remains so that the TSE (and by consequence Φ-analysis) is still a meaningful concept for Villin.

In order to obtain a complete picture of the folding kinetics for a protein, the observation of a large number of folding trajectories is crucial. This might be particularly important for protein A, given that the inconsistencies between various computational studies for this protein may lie in the existence of multiple transition states and pathways (36). We recently developed a minimalist transferable all-atom model (37), which was successfully used to predict *ab initio* the native structures of a diverse set of proteins including α, β, α+β, and α/β proteins. Here we apply the same model to protein A and a single-point mutant of Villin, but go beyond structure prediction by carefully exploring their folding behavior as an ensemble process. Moreover, we determine the TSEs for two proteins. To achieve this, we make use of $p_{fold}$ analysis, which requires additional simulations but is the most reliable method for identifying the TSE (38, 39). Combining the results obtained here with previous results for the Engrailed Homeodomain (ENH) (40) allows us to formulate a universal framework for the folding of small 3-helix bundle proteins within which we find a substantial diversity in the details of the folding mechanism.

## Results

We perform 2000 MC dynamics folding trajectories for protein A and Villin at a single temperature ($T \approx 300$ K), starting from random initial conformations (see Supporting Text). The large number of folding trajectories and the long total simulation time ($\approx 70$ ms) can be achieved because of the relative simplicity of our transferable all-atom protein model (Eq. **1** in Methods). Our objective is, based on these simulation results, to identify and compare robust features of the folding mechanism for the two proteins.

**Initial selection of trajectories.** Not all of the 2000 trajectories contain native-like low-energy structures. Therefore, before turning to the folding kinetics, we make an initial objective selection of a set of "representative" trajectories that fold into native-like conformations. This selection of trajectories is based on a simple clustering procedure of the lowest-energy structures obtained for the respective trajectories (this procedure is different from the structural kinetic cluster analysis performed on the full trajectories below). Hence, we first collect for each trajectory the lowest-energy structure observed in that trajectory. This set of 2000 conformations (each one representing a trajectory) is then clustered using their pairwise root-mean-square deviation, RMSD, into a simple single-link graph. In this graph, each node represents a conformation and edges are drawn between any two conformations (nodes) whose RMSD is below a threshold value $d_c$. Finally, we select the cluster with the highest average connectivity, $<k>$, where $k$ is the number of edges of a node and $<>$ is the cluster average. With our choice of $d_c$ (1.1 and 1.5 Å for protein A and Villin, respectively) the two clusters selected for protein A and Villin contain roughly the same number of structures (147 and 149 for protein A and Villin, respectively). All structures within these clusters are highly similar to each other and, more importantly, they are structurally highly similar to the respective experimental structures (<RMSD> = 2.7 and 2.8 Å for protein A and Villin, respectively).

Having made this selection of lowest-energy structures based on their structural connectivity, we will in what follows focus on the corresponding 147 and 149 trajectories respectively which are now guaranteed to proceed to low-energy, highly native-like states. A selection criterion based on structural connectivity among minimum-energy structures is objective as it does not require the knowledge of the native conformation. It is also relatively robust with respect to the choice of the cutoff value $d_c$, although nonnative-like clusters can sometimes have comparable connectivities $<k>$ (see Fig. S1 in Supporting Information). In particular, we find that clusters representing the "mirror image" topologies of the helix bundles are highly connected. A similar in spirit connectivity criterion has been used previously to identify high-quality candidates in protein structure prediction contexts (41). We find that the center structures within each of the two selected clusters for protein A and Villin, i.e. the top-$k$ structures, are indeed at least as native-like as the cluster average (see Fig. 1).

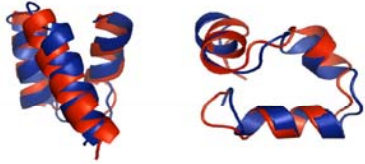

**Figure1.** *Comparison between the native structures (in blue) and superimposed minimum-energy top-k structures (in red) obtained through a clustering procedure for protein A (left panel) and Villin (right panel), as described in the text. The RMSD values between top-k and experimental structures are 2.7 and 2.1 Å for protein A and Villin, respectively. Structures are displayed by using PyMOL (42).*

**Chain collapse vs secondary structure formation.** We begin by examining the relaxation behavior of the two proteins. Fig. 2 compares the chain collapse and helix formation as obtained from the selected trajectories. A common property for the two proteins is a relatively rapid initial collapse of the chain, although it is slightly faster for Villin. Due to this fast "burst" phase, we find that the $R_g$ relaxation is well described by a double-exponential function for both protein A and Villin (see Fig. 2, upper panels). Similar fits are obtained for the total energy $E$ and the RMSD with similar fit parameters (see Fig. S2). We find that the two time constants are separated roughly by an order of magnitude and we associate the slowest relaxation phase (time constant $\tau_{slow}$) with the overall folding process. Averaging over the three observables ($R_g$, $E$, and, RMSD), we find $<\tau_{slow}> = 43.0 \times 10^6$ MC steps and $<\tau_{slow}> = 42.8 \times 10^6$ MC steps for protein A and Villin, respectively, so that protein A and Villin fold approximately at the same rate in our model. This result is in rough agreement with the corresponding experimental time constants which are 8.6 $\mu$s at 310 K (extrapolated) for protein A (7) and 5 $\mu$s at 300 K for Villin (6).

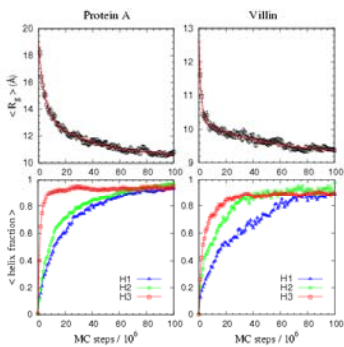

**Figure 2.** *Relaxation behavior of the average radius of gyration, $R_g$ (top panels) and average fraction helicity (lower panles) of each individual helix, obtained at $T \approx 300$ K. Helicity is determined using the criterion of Kabsch and Sander (43). The $R_g$ relaxation data are fitted (red curves, top panels) to a double-exponential function,* $f(t) = a_1 \exp(-t/\tau_{fast}) + a_2 \exp(-t/\tau_{slow}) + b$, *using a Levenberg-Marquardt fit procedure with $a_1$, $a_2$, $\tau_{fast}$, $\tau_{slow}$, and b as free parameters.*

While the collapse behavior is similar for the two proteins, we observe differences in the way secondary structure is formed, as can be seen from Fig. 2 (lower panels). For protein A, the initial collapse phase coincides with the formation of H3. H1 and H2 are also formed during the collapse, albeit only partially. Hence, there is a substantial overall coil-to-helix transition in the initial phase of folding. In contrast to protein A, Villin exhibits relatively fast formation of both H2 and H3 during the initial chain collapse while H1 forms at a slower rate. Although helix formation can be fast in our model, as exemplified by H3 in protein A, we find that it is not unrealistically fast however. Laser-induced *T*-jump experiments have been obtained for protein A using IR spectroscopy (7) and for the Villin headpiece using tryptophan flourescence (6). In both studies, a fast phase ($\approx \tau_{slow}/100$) was detected and interpreted to be related to fast helix melting and formation. It is important to note that these studies are $T$-jump unfolding experiments and, as such, cannot be directly related our folding kinetics results. However, they show that it is possible for helix formation to occur on very fast time scales relative to the overall folding transition even for extremely fast folding proteins like protein A and Villin.

**Structural kinetic cluster analysis.** Although the time-dependence of secondary structure formation and chain collapse in Fig. 2 give useful information, this type of analysis does not provide details about structural states during the folding process. We therefore turn to a structural cluster procedure developed by our group (40). The basic idea is centered around the concept of a "structural graph" (for an extensive discussion see Ref. (40) and also Fig. S3) which aims to provide structural and kinetic information about coarse-grained features of the folding process. The structural graph is created in two steps. In the first step, all snapshots from all trajectories (147 and 149 for protein A and Villin, respetively) are treated on an equal footing and clustered together into a single-link graph; this approach sets it apart from some other cluster procedures used to analyze folding trajectories (44, 45). Each conformation is represented by a node and two nodes are linked by an edge if their structural similarity $d$ is less than a cutoff, $d_c$. We determine $d_c$ based on the total number of conformations in the Giant Component (GC), i.e. the largest cluster, which represents the native basin of attraction N (for suitable $d$'s). In the second step, information about the trajectories is re-introduced in order to kinetically characterize the clusters. A key quantity is the *flux*, $F$, defined as the fraction of all trajectories passing through the cluster. Hence, clusters through which all trajectories pass have $F = 1$. This is the situation for the GC, for example, as every trajectory eventually reaches N. Other clusters with $F = 1$ can be interpreted as obligatory intermediate states (40). In addition to $F$, we calculate for each cluster the mean-first-passage-time, MFPT, and the mean-least-exit-time, MLET. The $F$, MFPT, and MLET quantities along with the structural characteristics of the obtained clusters provide a powerful yet simple way of understanding details about the folding process from an ensemble perspective.

The clustering of the snapshots can in principle be performed using any structural similarly measure $d$. Here we follow Hubner et al. (40) and construct structural graphs using the three order parameters RMSD, DRMS, and $\Delta R_g$. Because of the different characteristics of the parameters, each one provides a different perspective on the folding process. Note that the cluster properties we focus on here are "coarse-grained" in nature, which is necessary in a MC study where the dynamics at very short time scales may depend on chain update properties. At longer time scales, however, the detailed balance criterion guarantees that averaged properties will become increasingly accurate.

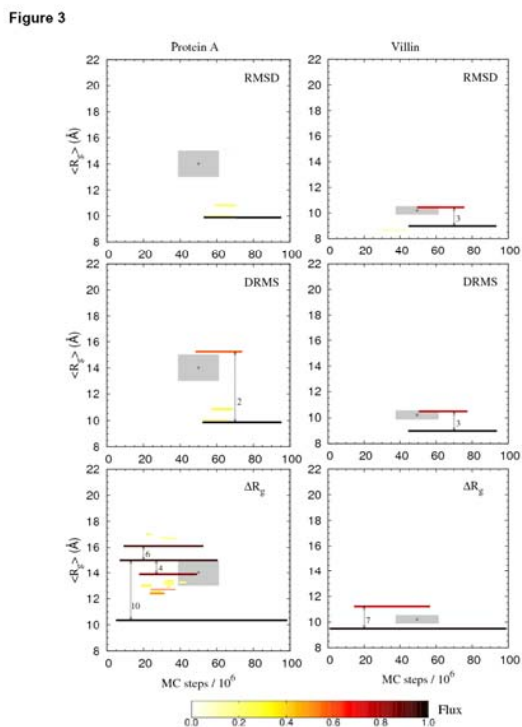

**Figure. 3.** *Results of the structural kinetic cluster analysis for protein A and Villin. Each cluster is represented by a line, from t = MFPT to t = MLET, and color-coded by its flux, F as indicated by the color scale. Only clusters with F > 0.1 are shown. The vertical location of each cluster is determined by the average radius of gyration, <$R_g$>, where <> is the cluster average. The location of the two TSEs for protein A and Villin are indicated by shaded areas*

centered (+) around the average $R_g$ and time $t$, with averages taken over the two ensembles; the sizes of the two shaded areas reflect $1\sigma$ deviations in both the $R_g$ and $t$ directions. The RMSD structural graphs are obtained using the cutoffs $d_c$ = 1.1 Å and 1.5 Å for protein A and Villin, respectively, while for DRMS we use $d_c$ = 0.9 Å and 1.2 Å, respectively. With these choices of $d_c$, the RMSD and DRMS GCs contains ≈ 35% of all conformations, a reasonable number given that the trajectory time is ≈ $2\tau_f$. For $\Delta R_g$, all reasonable choices of $d_c$ give larger GCs than for RMSD and DRMS, indicating that it contains not only native-like structures (see text). The results for $\Delta R_g$ are not very sensitive to the specific choice of $d_c$. Results are shown obtained for $d_c$ = 0.0080 Å and 0.0040 Å, respectively.

Figure 3 shows the results of our structural graph analysis for the two proteins. Clusters are represented as horizontal lines color-coded according to their flux $F$ and drawn from $t$ = MFPT to MLET. Starting with protein A, we find the RMSD and DRMS structural graphs to be dominated by a single high-flux cluster (i.e. the GC) which we associate with the native state. The absence of clusters at early times $t$ during the folding process ($t < \tau_f$) means that during pre-transition state folding times no accumulation of structurally similar conformations occurs. The $\Delta R_g$-structural graph provide a somewhat different perspective by reporting mainly on chain size and, by contrast, exhibits several early high-flux clusters, which in fact have $F \approx 1$. This means that during early folding times, although the RMSD and DRMS graph exclude the possibility of significant populations of structurally coherent states, all trajectories fluctuate widely in size. In fact, transitions between the early clusters in the $\Delta R_g$ structural graph are numerous, approximately 5-10 per trajectory (see Fig. 3). In the $\Delta R_g$ structural graph, we see that the GC is a low-$R_g$ cluster with MFPT ≈ 4×10$^6$ MC steps << $\tau_f$. As opposed to the RMSD and DRMS structural graphs, where the GC represents the native state, the GC in the $R_g$ graph must therefore contain not only conformations that are part of the native basin of attraction but also pre-TSE compact conformations. A key question is then whether the native state is reached through a path within this low-$R_g$ GC cluster, or through another path involving more extended conformations. To answer this question, it is useful to consider the location of the transition-state ensemble (TSE) which is shown as a shaded area in Fig. 3. The determination of the TSE is discussed in detail below. From the somewhat extended nature of the TSE, we see that it is highly unlikely that N is reached by remaining in the low-$R_g$ GC cluster, i.e. through a series of compact states. Instead, the TSE is located during fluctuations to more extended conformations after which the chain collapses into the native state. In this sense, the early low-$R_g$ states are "off-pathway" (a similar behavior was found for another three-helix bundle protein, ENH (40)). For the Villin headpiece we find, as for protein A, that the largest fluctuations in chain size occur during early times in the folding process. However, after the initial collapse phase, the Villin chain remains fairly compact throughout the rest of the folding process which is clear from the $\Delta R_g$ structural graph in Fig. 3. This is also consistent with the TSE obtained for Villin which is relatively compact, as shown below.

Finally, we note that both protein A and Villin exhibit a semi-high flux cluster ($F \approx 0.6\text{-}0.7$) in the later stages of the folding process (see Fig. 3), and their structural properties and their role in the folding process appear to be remarkably similar. Both clusters are overlapping in time with the native basin of attraction. A closer analysis of the DRMS structural graph for Villin reveals that for 61% of the conformations in this late semi-high $F$ cluster, the corresponding trajectories have passed previously into the GC. For protein A, the corresponding fraction is 71%. From this perspective, these two clusters can be characterized as non-obligatory "post-TSE" intermediate states (similar to "hidden intermediates" found recently (33)). On the other hand, we find also that there is a small but significant overlap between the TSEs and these late intermediates: 9 out of 46 and 16 out of 57 TSE structures, respectively, are present in the protein A and Villin intermediates.

Structurally, too, it is evident that the two intermediates resemble a small part of the "conformational space" of the TSE. Both intermediates are characterized by a well-formed H3 detached from a native-like H1-H2 segment, which fits into the overall pattern of the TSEs (see below). It should be pointed out, however, that the two intermediate states are much more structurally coherent than the TSEs. Clearly then, this means that the stability of conformations within our transition states is not uniform making the distinction between transition state and intermediate a nontrivial issue. Transition states with some degree of polarization have been noted previously in the src SH3 domain (46). Delineating this intriguing issue is important in particular in the context of simple protein models but it is beyond the scope of the present work. However, a kinetic observable such as $p_{fold}$ (see below) is clearly necessary in order to examine the issue. For now, we simply note that the two intermediates detected here are non-obligatory states which partially overlap with the respective TSEs.

**Transition state ensembles.** The transition state is key to understanding the folding process as it defines the rate-limiting step for folding. We construct the TSEs directly through the $p_{fold}$ analysis, which is a natural and highly reliable way of determining the TSE. This analysis is based on the notion that each conformation in the TSE has a unique property, namely that trajectories starting from such a conformation have an equal chance of first reaching the native state and the unfolded state, given random initial conditions. We make use of this definition in finding the "true" TSE for our two proteins by first identifying a set of putative transition-state structures, and then confirming or rejecting them based on the probability of folding, $p_{fold}$, obtained by additional simulations. Since the RMSD GC cluster corresponds to the native state, we hypothesize that most viable putative transition-state structures can be found by selecting structures that immediately precede entry into the GC in the structural graph. This gives us a set of 783 and 798 putative transition-state structures for protein A and Villin, respectively. For each conformation in these putative sets, 100 independent trajectories are initiated randomly and conformations with 0.4 < $p_{fold}$ < 0.6 are taken be part of the TSE (see Supporting Information). This procedure generates a set of 46 and 57 "true" transition-state structures for protein A and Villin, respectively. The structures are illustrated in Fig. S4 and the coordinates are published at http://www-shakh.harvard.edu/db/TSE.html.

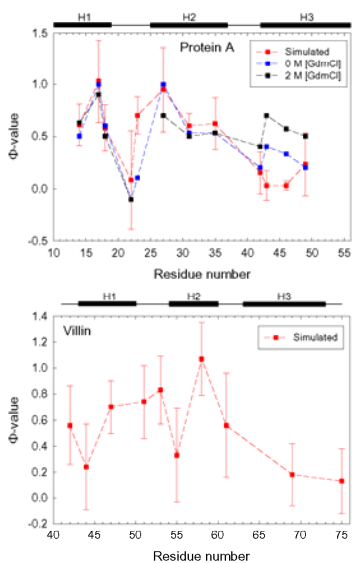

**Figure. 4.** *Comparison between simulated and experimental (32) Φ-values. At positions where more than one experimental Φ-values are reported, the average value is used and their individual experimental Φ-values on different types of mutations are shown in Fig. S5 in Supporting Information. Error bars denote the standard deviation, σ.*

Having obtained a representative sample of the transition state using the stringent $p_{fold}$ criterion, we use this set of structures to calculate theoretical Φ-values for our two proteins. We follow previous convention and interpret $\Phi_i$ for a residue $i$ as the number of contacts present in the TSE for residue $i$ divided by the number of native contacts (in $i$) (Eq. **S3**). We included all $\Phi^{sim}$ values with standard deviation, $\sigma < 0.5$ and the result of our $\Phi$-value calculations is given in Fig. 4. Experimental $\Phi$-values have been obtained previously for protein A (31, 32) and the agreement between theory and experiment is excellent, with an average absolute deviation ($|\Phi^{exp} - \Phi^{sim}|$) of 0.16 taken over all data points. We find that the most structured regions in the TSE are H1 and H2, as indicated by high $\Phi$-values, which form a native-like helical hairpin, while H3 is only weakly interacting with H1 and H2. This picture is in good agreement with two other detailed all-atom studies (12, 21), but not with results from a simpler Go-type study (9) which indicated initial H2-H3 formation rather. Of the obtained $\Phi^{sim}$ values only one differs significantly from the experimental value, namely $\Phi^{sim}_{L23} > \Phi^{exp}_{L23}$, located in the H1-H2 turn region. Our model thus predicts a highly ordered H1-H2 turn region, in the TSE. This may be related to the use of implicit water in our model. With implicit water, unsatisfied intramolecular hydrogen bonds cannot be compensated for by hydrogen bonding to water molecules which is likely to happen in a poorly ordered H1-H2 helix loop. We also underestimate somewhat $\Phi$-values in H3, which may indicate that H3 participates less in the TSE than suggested by experiments. For all other positions, a good *quantitative* agreement between $\Phi^{sim}$ and $\Phi^{exp}$ exists, which has been challenging to obtain in previous simulation studies (31).

For Villin, only one $\Phi$-value has been published so far, $\Phi_{K65} \approx 1.3$ in H3, which was obtained by a Lysine to Norleucine mutation designed to speed up folding (34). It is unclear to what extent this somewhat unconventional mutation can be interpreted in standard $\Phi$-value language but it appears to suggest that the N-terminal side of H3 is highly involved in the TSE. Although we were unable to obtain a $\Phi$-value at position K65, our results indicate by contrast that the most organized region of the TSE is centered around the H1-H2 segment. Overall, the situation is therefore similar to that of protein A. However, we note that TSE is overall less organized for Villin (Fig. S4). Also, we find quite large variations in the Φ-values within both H1 and H2. Low Φ-values are observed in the N-terminal ends of both H1 and H2.

## Discussion

Using a relative simple transferable sequence-based all-atom model, we performed a large number of *ab initio* protein folding runs for protein A and Villin headpiece providing us with necessary data to study the folding kinetics as an ensemble process. By combining our results, we obtain a "coarse-grained" picture of the folding processes of two 3-helix bundle proteins. Qualitatively the folding scenarios are similar for both proteins, and for another 3-helix bundle protein, ENH (40). For protein A, the initial collapse phase coincides with the formation of H3 and a partial formation of H1 and H2. In the subsequent slower phase, H1 and H2 continue to form while the chain visits both compact and non-compact states. Although there is significant secondary structure, these states lack specific global structural characteristics as we see from the absence of high-flux RMSD and DRMS early structural clusters (Fig. 3). Only when H1 and H2 are sufficiently structured can the transition state be reached, in which the H1-H2 segment is native-like forming a relatively ordered helical hairpin. H3, by contrast, has only limited interaction with this H1-H2 "nucleus" of the transition state. The folding process for Villin differs from protein A's in some aspects. The initial collapse in Villin is accompanied by the formation of both H2 and H3, while H1 forms at a slower rate. We also find that the Villin chain remains mostly compact during the remaining part of the folding process, as shown by the cluster analysis which exhibits only low-$R_g$ clusters. The TSE of Villin is characterized by relatively well-formed secondary elements and a native-like H1-H2 segment, similar to the situation in protein A but the TSE is structurally less coherent. The Villin TSE is also quite compact although it is slightly more extended than both the native structure and the early disordered compact states which follow the initial collapse.

The chain collapse behavior in the initial phase of folding which we find in our simulations have been observed, although with some variations, in several other simulation studies of both protein A (11, 12, 14, 16, 17, 40) and Villin (16, 17, 25-27). For protein A, this type of behavior may at first glance seem inconsistent with experiments which generally indicate that protein A folds in apparent two-state kinetics (31, 32, 47). Two observations most often used to support two-state kinetics is single-exponential relaxation and a linear dependence of ln $k_f$ on denaturant concentration [D], where $k_f$ is the folding rate, i.e. V-shaped Chevron plots. However, the folding relaxation kinetics for protein A have only been resolved for times $t > 150$ $\mu$s (31) while our collapse phase occurs on a much faster time scale. Hence, the fast collapse phase we observe would only be detected if much higher time resolution folding experiments can be obtained for protein A. Highly time-resolved experiments has been performed on apomyoglobin, where folding was initiated from a cold-denatured state, and a fast initial collapse phase was detected and found to be completed within 7$\mu$s (48). Linear refolding arms in the Chevron plots of protein A and various mutants have been observed in several studies (31, 32, 47, 49) but the folding rate $k_f$ can usually not be determined at very low [D] which might be important. A unique insight into the folding of protein A was obtained in a recent single-molecule study of protein A using FRET (49). Although two distinct populations was observed in the FRET efficiency (showing that U and N are distinct populations) a shift of the peak corresponding to U was observed with varying [D] indicating a chain compaction of U as [D] decreases. Whether this compaction of U would result in Chevron rollovers for [D] approaching 0, as observed when a collapsed state is artificially stabilized (50), remains to be seen. Regardless of this, we find that our results are in good agreement with the bimodal distribution of the FRET signal (see Fig. 6 in Ref.(50)). This can be seen from a clear separation in the probability distributions of $1/r^6_{ee}$ (mimicking FRET efficiency) for conformations in the U and N states, respectively, where $r_{ee}$ is the chain end-to-end distance (Fig. S6). Finally, we note that direct evidence for a compact unfolded state with high degree of nonnative hydrophobic interactions has been recently observed in the Trp-cage miniprotein TC5b using a novel type of NMR pulse-labeling experiment (51).

In terms of achieving a good agreement between theory and experiment, the excellent correspondence we find between $\Phi^{exp}$ and $\Phi^{sim}$ obtained for protein A is encouraging and means that, overall, the characteristics of the TSE is in very good agreement with the $\Phi$-analysis performed by Fersht *et al.* (31, 32) across the entire chain. A remaining issue is the extent to which secondary structure is present in the TSE. Our results indicate the presence of significant amounts of helicity in all three helices. This includes H3 which scores low $\Phi^{sim}$-values mainly because of its weak interaction with H1 and H2, in the TSE. Our finding that H3 is structured in the TSE appears to be at odds with the conclusion made by Fersht et al from their Ala→Gly scanning study (31, 32). $\Phi$-values from such mutations, when performed at protein surface positions, were interpreted as mainly probing the secondary structure content of the TSE because no tertiary contacts are deleted upon mutation (52). In light of these experiments, it is therefore possible that the stability of secondary structure elements, in particular H3, are somewhat overestimated in our model. However, we note that low Ala→Gly $\Phi^{exp}$-values in H3 could also be explained by residual H3 secondary structure in the denatured state D under folding conditions. This situation would produce small $\Delta\Delta G_{D\text{-}\ddagger}$'s upon mutation and consequently small $\Phi$-values. There are two factors that indicate that this might indeed be the case. First, all Ala→Gly $\Phi^{exp}$-values in H3 are markedly larger at 2M GdmCl than at 0M GdmCl (see Table 1 in Ref. (32)), which is consistent with the melting of residual secondary structure in the denatured state at 2M GdmCl compared to 0M GdmCl. Importantly, this trend does not exist for H1 or H2. Second, H3 is the only one of the helices which exhibits some stability on its own. i.e. as an individual fragment (33). Hence, it appears likely that some residual secondary structure exists in H3 in the denatured state D under folding conditions which may be an alternative explanation for the low Ala→Gly $\Phi^{exp}$-values in H3.

## Conclusions and Outlook

We have demonstrated that a simple and computationally tractable transferable all-atom model can capture details of the folding behavior of two small helical proteins at a quantitative level. In particular, we find that the obtained $\Phi$-values for protein A fit experimental data to a degree that has not been achieved by previous simulation studies, while future experiments will have to be conducted to test the validity of the obtained $\Phi$-values for the Villin headpiece.

This study along with a previous investigation of the ENH (40) provides a comprehensive anlysis of folding processes for three-helix bundle proteins at an atomistically detailed level. When we combine the results from these studies a univeral picture of the folding of three-helix bundle proteins emerges. The first step is an initial collapse of the chain accompanied by partial formation of the α-helices (to a greater or lesser extent). On average, the chain remains relatively compact, but frequent visits to more extended structures occur. During such fluctuations, the TSE can be located after which the chain collapses to N. The TSE consists of relatively well-formed helices organized into a two-helix hairpin and a third helix which is partially detached. Within this general framework there can be significant differences in the details, however. For example, the initial collapse phase can be accompanied by the formation of a single helix (such as H3 for protein A) or two helices (H2 and H3 for Villin). Moreover, there are two possibilities for the helical hairpin in the TSE, which is dominated by a H1-H2 in both protein A and Villin. Our simulations (40) and recent experimental work of Fersht and coworkers found that H2-H3 hairpin in ENH forms an independently stable domain (53). Our analysis suggests that formation of a helix-turn-helix motif prior to entering the TSE is perhaps a universal mechanism observed in folding of 3-helix-bundle proteins, although details of which hairpin is formed may vary.

It would, of course, be highly interesting to examine the folding behavior of proteins with other classes of protein folds with a similar approach. A crucial question is to what extent the universal features of 3-helix-bundle folding discovered here applies to more complicated folds. Extending the present work to the folding processes a set of proteins involving β-sheet structure is natural, especially in the light of previous successes of the present model to treat a diverse set of proteins within the context of structure prediction (37).

## Methods

**Energy function.** The all-atom energy function $E$ in our previous study (37) has been further developed and now takes the form:

$$E = E_{con} + w_{trp} \times E_{trp} + w_{hb} \times E_{hb} + w_{sct} \times E_{sct} \qquad [1]$$

where $E_{con}$ is the pairwise atom-atom contact potential, $E_{hb}$ is the hydrogen bonding potential, $E_{trp}$ is the sequence-dependent local torsional potential based on the statistics of sequential amino acid triplets, and $E_{sct}$ is the side-chain torsional angle potential (see Supporting Text). Detailed information on the first three energy terms can be found in our previous publication (37). It should be noted that secondary structure information from PSIPRED is not used in this study, which enables us to observe true *ab initio* folding of proteins (see Supporting Text).

# Acknowledgements

This work is supported by the NIH. We thank Dr. Chaok Seok for help with the implementation of the local move set that conserves detailed balance.

# References

1.  Alm, E. & Baker, D. (1999) Matching theory and experiments in protein folding. *Curr. Opin. Struct. Biol.* **9,** 189-196.
2.  Gianni, S., Guydosh, N. R., Khan, F., Caldas, T. D., Mayor, U., White, G. W., DeMarco, M. L., Daggett, V., & Fersht, A. R. (2003) Unifying features in protein-folding mechanisms. *Proc. Natl. Acad. Sci. USA* **100,** 13286-13291.
3.  Laurents, D. V. & Baldwin, R. L. (1998) Protein folding: Matching theory and experiment. *Biophys. J.* **75,** 428-434.
4.  Pande, V. S. (2003) Meeting halfway on the bridge between protein folding theory and experiment. *Proc. Natl. Acad. Sci. USA* **100,** 3555-3556.
5.  Lo Conte, L., Ailey, B., Hubbard, T. J., Brenner, S. E., Murzin, A. G., & Chothia, C. (2000) SCOP: a structural classification of proteins database *Nucleic Acids Res.* **28,** 257-259.
6.  Kubelka, J., Eaton, W. A., & Hofrichter, J. (2003) Experimental tests of villin subdomain folding simulations. *J. Mol. Biol.* **329,** 625-630.
7.  Vu, D. M., Myers, J. K., Oas, T. G., & Dyer, R. B. (2004) Probing the folding and unfolding dynamics of secondary and tertiary structures in a three-helix-bundle protein. *Biochemistry* **43,** 3582-3589.
8.  Alonso, D. O. V. & Daggett, V. (2000) Staphylococcal protein A: Unfolding pathways, unfolded states, and differences betwen the B and E domains. *Proc. Natl. Acad. Sci. USA* **97,** 133-138.
9.  Berriz, G. F. & Shakhnovich, E. I. (2001) Characterization of the folding kinetics of a three-helix bundle protein *via* a minimalist Langevin model. *J. Mol. Biol.* **310,** 673-685.
10. Cheng, S., Yang, Y., Wang, W., & Liu, H. (2005) Transition state ensemble for the folding of B domain of protein A: A comparison of distributed molecular dynamics simulations with experiments. *J. Phys. Chem. B* **109,** 23645-23654.
11. Favrin, G., Irbäck, A., & Wallin, S. (2002) Folding of a small helical protein using hydrogen bonds and hydrophobicity forces. *Proteins* **47,** 99-105.
12. García, A. E. & Onuchic, J. N. (2003) Folding a protein in a computer: An atomic description of the folding/unfolding of protein A. *Proc. Natl. Acad. Sci. USA* **100,** 13898-13903.
13. Ghosh, A., Elber, R., & Scheraga, H. A. (2002) An atomically detailed study of the folding pathways of protein A with the stochastic difference equation. *Proc. Natl. Acad. Sci. USA* **99,** 10394-10398.
14. Guo, Z., Brooks III, C. L., & Boczko, E. M. (1997) Exploring the folding free energy surface of a three-helix bundle protein. *Proc. Natl. Acad. Sci. USA* **94,** 10161-10166.
15. Hubner, I. A., Deeds, E. J., & Shakhnovich, E. I. (2005) High-resolution protein folding with a transferable potential. *Proc. Natl. Acad. Sci. USA* **102,** 18914-18919.
16. Jang, S., Kim, E., Shin, S., & Pak, Y. (2003) Ab initio folding of helix bundle proteins using molecular dynamics simulations. *J. Am. Chem. Soc.* **125,** 14841-14846.
17. Kim, S. Y., Lee, J., & Lee, J. (2004) Folding of small proteins using a single continuous potential. *J. Chem. Phys.* **120,** 8271-8276.
18. Kussell, E., Shimada, J., & Shakhnovich, E. I. (2002) A structure-based method for derivation of all-atom potentials for protein folding. *Proc. Natl. Acad. Sci. USA* **99,** 5343-5348.
19. Linhananta, A. & Zhou, Y. (2002) The role of sidechain packing and native contact interactions in folding: Discontinuous molecular dynamics folding simulations of an all-atom Gō model of fragment B of *Staphylococcal* protein A. *J. Chem. Phys.* **117,** 8983-8995.

20. Zhou, Y. & Karplus, M. (1999) Interpreting the folding kinetics of helical proteins. *Nature* **401,** 400-403.

21. Boczko, E. M. & Brooks, C. L., 3rd (1995) First-principles calculation of the folding free energy of a three-helix bundle protein *Science* **269,** 393-396.

22. De Mori, G. M., Colombo, G., & Micheletti, C. (2005) Study of the Villin headpiece folding dynamics by combining coarse-grained Monte Carlo evolution and all-atom Molecular Dynamics. *Proteins* **58,** 459-471.

23. Herges, T. & Wenzel, W. (2005) Free-energy landscape of the Villin headpiece in an all-atom force field. *Structure* **13,** 661-668.

24. Lei, H., Wu, C., Liu, H., & Duan, Y. (2007) Folding free-energy landscape of villin headpiece subdomain from molecular dynamics simulations. *Proc. Natl. Acad. Sci. USA* **104,** 4925-4930.

25. Zagrovic, B., Snow, C. D., Khaliq, S., Shirts, M. R., & Pande, V. S. (2002) Native-like mean structure in the unfolded ensemble of small proteins. *J. Mol. Biol.* **323,** 153-164.

26. Zagrovic, B., Snow, C. D., Shirts, M. R., & Pande, V. S. (2002) Simulation of folding of a small alpha-helical protein in atomistic detail using worldwide-distributed computing. *J. Mol. Biol.* **323,** 927-937.

27. Fernandez, A., Shen, M. Y., Colubri, A., Sosnick, T. R., Berry, R. S., & Freed, K. F. (2003) Large-scale context in protein folding: *Biochemistry* **42,** 664-671.

28. Duan, Y. & Kollman, P. A. (1998) Pathways to a protein folding intermediate observed in a 1-microsecond simulation in aqueous solution. *Science* **282,** 740-744.

29. Kinnear, B. S., Jarrold, M. F., & Hansmann, U. H. (2004) All-atom generalized-ensemble simulations of small proteins *J. Mol. Graph. Model.* **22,** 397-403.

30. Wolynes, P. G. (2004) Latest folding game results: Protein A barely frustrates computationalists. *Proc. Natl. Acad. Sci. USA* **101,** 6837-6838.

31. Sato, S., Religa, T. L., & Fersht, A. R. (2006) Φ-Analysis of the folding of the B domain of Protein A using multiple optical probes. *J. Mol. Biol.* **360,** 850-864.

32. Sato, S., Religa, T. L., Daggett, V., & Fersht, A. R. (2004) Testing protein-folding simulations by experiments: B domain of protein A. *Proc. Natl. Acad. Sci. USA* **101,** 6952-6956.

33. Bai, Y., Karimi, A., Dyson, H. J., & Wright, P. E. (1997) Absence of a stable intermediate on the folding pathway of protein A. *Protein Sci.* **6,** 1449--1457.

34. Chiu, T. K., Kubelka, J., Herbst-Irmer, R., Eaton, W. A., Hofrichter, J., & Davies, D. R. (2005) High-resolution x-ray crystal structures of the villin headpiece subdomain, an ultrafast folding protein. *Proc. Natl. Acad. Sci. USA* **102,** 7517-7522.

35. Kubelka, J., Chiu, T. K., Davies, D. R., Eaton, W. A., & Hofrichter, J. (2006) Sub-microsecond protein folding. *J. Mol. Biol.* **359,** 546-553.

36. Itoh, K. & Sasai, M. (2006) Flexibly varying folding mechanism of a nearly symmetrical protein: B domain of Protein A. *Proc. Natl. Acad. Sci. USA* **103,** 7298--7303.

37. Yang, J. S., Chen, W. W., Skolnick, J., & Shakhnovich, E. I. (2007) All-atom *ab initio* folding of a diverse set of proteins. *Structure* **15,** 53-63.

38. Snow, C. D., Rhee, Y. M., & Pande, V. S. (2006) Kinetic definition of protein folding state ensembles and reaction coordinates. *Biophys. J.* **91,** 14-24.

39. Du, R., Pande, V. S., Grosberg, A. Y., Tanaka, T., & Shakhnovich, E. I. (1998) On the transition coordinate for protein folding. *J.Chem. Phys.* **108,** 334-350.

40. Hubner, I. A., Deeds, E. J., & Shakhnovich, E. I. (2006) Understanding ensemble protein

folding at atomic detail. *Proc. Natl. Acad. Sci. USA* **103,** 17747-17752.

41.     Shortle, D., Simons, K. T., & Baker, D. (1998) Clustering of low-energy conformations near the native structures of small proteins. *Proc. Natl. Acad. Sci. USA* **95,** 11158-11162.

42.     DeLano, W. L. (2002) *The PYMOL Molecular Graphics System*, (DeLano Sci., San Carlos, CA).

43.     Kabsch, W. & Sander, C. (1983) Dictionary of protein secondary structure: pattern recognition of hydrogen-bonded geometrical constraints. *Biopolymers* **22,** 2577-2637.

44.     Karpen, M. E., Tobias, D. J., & Brooks, C. L., 3rd (1993) Statistical clustering techniques for the analysis of long molecular dynamics trajectories: analysis of 2.2-ns trajectories of YPGDV *Biochemistry* **32,** 412-420.

45.     Rao, F. & Caflisch, A. (2004) The protein folding network *J. Mol. Biol.* **342,** 299-306.

46.     Grantcharova, V. P., Riddle, D. S., Santiago, J. V., & Baker, D. (1998) Important role of hydrogen bonds in the structurally polarized transition state for folding of the src SH3 domain *Nat. Struct. Biol.* **5,** 714-720.

47.     Myers, J. K. & Oas, T. G. (2001) Preorganized secondary structure as an important determinant of fast protein folding. *Nat. Struct. Biol.* **8,** 552-558.

48.     Ballew, R. M., Sabelko, J., & Gruebele, M. (1996) Direct observation of fast protein folding: the initial collapse of apomyoglobin *Proc. Natl. Acad. Sci. U S A* **93,** 5759-5764.

49.     Huang, F., Sato, S., Sharpe, T. D., Ying, L., & Fersht, A. R. (2007) Distinguishing between cooperative and unimodal downhill protein folding *Proc. Natl. Acad. Sci. U S A* **104,** 123-127.

50.     Otzen, D. E. & Oliveberg, M. (1999) Salt-induced detour through compact regions of the protein folding landscape *Proc. Natl. Acad. Sci. U S A* **96,** 11746-11751.

51.     Mok, K. H., Kuhn, L. T., Goez, M., Day, I. J., Lin, J. C., Andersen, N. H., & Hore, P. J. (2007) A pre-existing hydrophobic collapse in the unfolded state of an ultrafast folding protein *Nature* **447,** 106-109.

52.     Scott, K. A., Alonso, D. O. V., Sato, S., Fersht, A. R., & Daggett, V. (2007) Conformational entropy of alanine versus glycine in protein denatured states. *Proc. Natl. Acad. Sci. USA* **104,** 2661-2666.

53.     Religa, T. L., Johnson, C. M., Vu, D. M., Brewer, S. H., Dyer, R. B., & Fersht, A. R. (2007) The helix-turn-helix motif as an ultrafast independently folding domain: the pathway of folding of Engrailed homeodomain *Proc. Natl. Acad. Sci. U S A* **104,** 9272-9297.

**Supporting Text and Figures**

**Simulation.** Protein A (1BDD) and Villin (1VII) are chosen as model proteins. Unstructured tails were truncated (res. 1-9; 57-60) for protein A, and W64 was mutated to A64 and a hydrophobic residue at the end, F76 was truncated for Villin. Starting from different random coil configurations, 2000 independent Monte Carlo simulation for each protein were conducted at $T = 0.60$ for $10^8$ steps. The melting temperature for protein A, $T_m^{\text{sim}} \approx 0.70$, was obtained from simulations at different temperatures (data not shown), while the experimental melting temperature is 346 K (1). The MC temperature $T = 0.60$ thus corresponds to $T \approx 300\text{K}$. Global and local moves were used for backbone rotation (2). To keep the detailed balance condition, a knowledge-based move (2, 3) was not used, and the local move set was modified (4, 5) (See below). For each trajectory, snapshots were stored at every $10^6$ MC step and the energy minimum structure was recorded at the end of simulation.

**The side-chain torsional angle energy.** The side-chain torsional angle energy (6) was obtained from the same database as our previous study (2) by

$$E_{\text{sct}} = \sum_i E_{A_{i-1}A_iA_{i+1}}, \quad E_{A_{i-1}A_iA_{i+1}} = \frac{-\mu N_j + (1-\mu)\widetilde{N}_j}{\mu N_j + (1-\mu)\widetilde{N}_j}, \qquad [S1]$$

where $N_j$ and $\widetilde{N}_j$ are the number of observations in the $j$-th bin of a side-chain torsional angle $\chi$'s of residue $A_i$ and total number of observances subtracted by $N_j$ for a triplet consisting of $A_{i-1}$, $A_i$, and $A_{i+1}$, respectively. The bin width was 30° and the value of $\mu = 0.991$ was chosen to make the net interaction zero.

**Detailed balance for the local move.** Dodd *et al.* (5) developed a local move which is composed of concerted rotation of seven adjacent bonds. They also showed that changes in these seven degrees of freedom are correlated and therefore a new sampling method rather than the conventional Metropolis rule should be used to conserve the detailed balance. We follow their procedure and probability of accepting a move from the old state $o$ to the new state $n$ is given by (5)

$$P(o \rightarrow n) = \min\left[1, \frac{N^{(n)} \exp(-U(n)/T) J(n)}{N^{(o)} \exp(-U(o)/T) J(o)}\right] \qquad \text{[S2]}$$

where $N$ is the number of solutions, $U$ is the potential energy, $T$ is temperature, and $J$ is the Jacobian determinant.

**$p_{\text{fold}}$ analysis.** Transition state ensemble and $\Phi$ values were obtained by the following procedure. From representative trajectories, structures that are just before entering the native cluster were selected as putative transition state structures. Starting from each of these structures, 100 short ($10^6$ MC steps) MC simulations were executed. If the trajectory finds a structure whose RMSD from the top-$k$ structure is less than 3 Å, then it is considered to be folded. It should be noted that the simulated $\Phi$-values are not sensitive to the choice of this criterion (see Fig. S7). A structure with $0.4 < p_{\text{fold}} < 0.6$ is regarded as a member of TSE, where $p_{\text{fold}}$ is the fraction of runs that folded out of 100 runs. The $\Phi$-values are calculated by

$$\Phi_i = \frac{\left\langle N_{\text{TSE}}^i \right\rangle}{\left\langle N_{\text{top-}k}^i \right\rangle} \qquad \text{[S3]}$$

where $\left\langle N_{\text{top-}k}^i \right\rangle$ and $\left\langle N_{\text{TSE}}^i \right\rangle$ are the average number of contacts at residue $i$ for 10 top-k structures and for TSE, respectively.

# References

1.    Vu, D. M., Myers, J. K., Oas, T. G. & Dyer, R. B. (2004) *Biochemistry* **43,** 3582-3589.

2.    Yang, J. S., Chen, W. W., Skolnick, J. & Shakhnovich, E. I. (2007) *Structure* **15,** 53-63.

3.    Chen, W. W., Yang, J. S. & Shakhnovich, E. I. (2007) *Proteins* **66,** 682-688.

4.    Coutsias, E. A., Seok, C., Jacobson, M. P. & Dill, K. A. (2004) *J. Comput. Chem.* **25,** 510-528.

5.    Dodd, L. R., Boone, T. D. & Theodorou, D. N. (1993) *Mol. Phys.* **78,** 961-996.

6.    Yang, J. S., Kutchukian, P. S. & Shakhnovich, E. I. *Submittted for publication*.

7.    Kabsch, W. & Sander, C. (1983) *Biopolymers* **22,** 2577-2637.

## Figure Legends

**Fig. S1.** Clustering results with various RMSD cutoffs, $d_c$ for (A) protein A and (B) Villin. Clusters which consist of more than 10 structures are shown for clarity. In all cases, structures in the cluster with the highest average connectivity, $<k>$ show very low average RMSD values, <RMSD> from the experimental structure. The cluster with the second highest $<k>$ is composed of mirror images.

**Fig. S2.** Relaxation behavior of the total energy RMSD($t$) and $E(t)$ for protein A and Villin, at $T \approx 300$ K. Fits to double-exponential curves (red curves) are good in all 4 cases. Relaxation times $\tau_{slow}$ and $\tau_{fast}$ for each curve are given in the plots.

**Fig. S3.** A schematic view explaining the concept of a structural graph and flux, $F$. Each node (colored ovals) represents a single conformation and edges (solid lines) are drawn between structurally similar conformations (as determined by the criterion $d < d_c$, where $d$ is a similarity measure). Different colors indicate conformations from different trajectories and wavy lines indicate the direction of time $t$. A collection of nodes that are linked by edges (either directly or in several steps) belong to the same cluster. To determine the kinetic significance of a cluster, we use the concept of flux, $F$, which is defined as the fraction of all trajectories that pass though a cluster. Therefore, since all conformation will fold into the native state, this cluster will have $F = 1$ (rightmost cluster). Other clusters with $F = 1$, through which all trajectories have to pass, can be interpreted as obligatory intermediate states.

**Fig. S4.** Structural superimpositions of all TSE structures obtained by $p_{fold}$ analysis for protein A (left) and Villin (right). The H1-H2 hairpin is formed in TSE for both proteins. The average total helicity calculated from the whole chain, as measured by the criterion of Kabsch and Sanders (7), is 81% and 65% for protein A and Villin, respectively. The average size of the ensembles, $<R_g>$, is 14.2 Å for protein A and 10.3 Å for Villin.

**Fig. S5.** Experimental Φ–values of different types of mutations for protein A. Our calculation method based on the number of contacts only in wide-type can be regarded as WT-G mutation. It is surprising to see that this method can reproduce experimental Φ–values for other types of mutations.

**Fig. S6.** Computational equivalence of FRET signals for protein A. The probability distribution of end-to-end distances, $1/r^6$, in the native and unfolded states, respectively. $r$ is determined as the $C_\alpha$-$C_\alpha$ distance between residues 10 and 56 (numbering as the 1BDD PDB entry). The native state is composed of the set of 147 lowest-$E$ conformations of the respective 147 trajectories used to study folding of protein A. The unfolded ensemble is composed of all snapshot taken at times $t < 20$ million MC steps excluding the initial configurations.

**Fig. S7.** Simulated Φ-values where the cutoff of 2 Å was used for being folded instead of 3 Å. The numbers of TSEs are 65 and 109 for protein A and Villin, respectively. Note that comparable results could be obtained regardless of the cutoff values. (For protein A, $|\Phi^{exp} - \Phi^{sim}| = 0.14$.)

# Figure S1

$d_c = 1.4$ Å

| ID* | $n$[†] | <rmsd>[‡] | <k>[§] |
|-----|-----|------|------|
| 1 | 173 | 2.74 | 42.7 |
| 2 | 388 | 8.37 | 37.9 |
| 3 | 366 | 6.11 | 21.9 |
| 4 | 85 | 9.53 | 9.7 |
| 5 | 130 | 12.09 | 8.6 |

$d_c = 2.0$ Å

| ID | $n$[†] | <rmsd>[‡] | <k>[§] |
|-----|-----|------|------|
| 1 | 268 | 3.30 | 29.1 |
| 2 | 127 | 9.43 | 24.6 |
| 3 | 90 | 6.79 | 20.3 |
| 4 | 650 | 7.14 | 12.1 |
| 5 | 184 | 6.73 | 8.0 |
| 6 | 28 | 4.96 | 4.9 |

$d_c = 1.1$ Å

| ID | $n$[†] | <rmsd>[‡] | <k>[§] |
|-----|-----|------|------|
| 1 | 147 | 2.68 | 12.2 |
| 2 | 298 | 8.31 | 11.5 |
| 3 | 134 | 5.42 | 8.9 |
| 4 | 43 | 11.82 | 3.6 |
| 5 | 28 | 9.79 | 3.6 |

$d_c = 1.5$ Å

| ID | $n$[†] | <rmsd>[‡] | <k>[§] |
|-----|-----|------|------|
| 1 | 149 | 2.82 | 11.8 |
| 2 | 96 | 9.35 | 10.5 |
| 3 | 54 | 6.67 | 7.6 |
| 4 | 31 | 5.97 | 6.5 |
| 5 | 68 | 5.48 | 5.3 |
| 6 | 37 | 7.88 | 4.9 |
| 7 | 76 | 7.96 | 4.7 |
| 8 | 25 | 6.81 | 3.1 |

$d_c = 0.9$ Å

| ID | $n$[†] | <rmsd>[‡] | <k>[§] |
|-----|-----|------|------|
| 1 | 74 | 2.63 | 4.4 |
| 2 | 102 | 7.99 | 4.0 |
| 3 | 39 | 5.30 | 2.7 |
| 4 | 23 | 8.08 | 2.4 |

*Cluster ID sorted by <k>.
[†]Number of structures in a cluster.
[‡]Average rmsd of structures from the experimental structure.
[§]Average connectivity of structures.

$d_c = 1.4$ Å

| ID | $n$[†] | <rmsd>[‡] | <k>[§] |
|-----|-----|------|------|
| 1 | 106 | 2.60 | 10.8 |
| 2 | 58 | 9.31 | 10.7 |
| 3 | 33 | 6.43 | 6.6 |
| 4 | 32 | 6.77 | 5.4 |
| 5 | 47 | 5.42 | 4.4 |
| 6 | 28 | 9.70 | 4.1 |
| 7 | 22 | 7.96 | 4.1 |
| 8 | 61 | 7.89 | 3.9 |

(A)

(B)

**Figure S2**

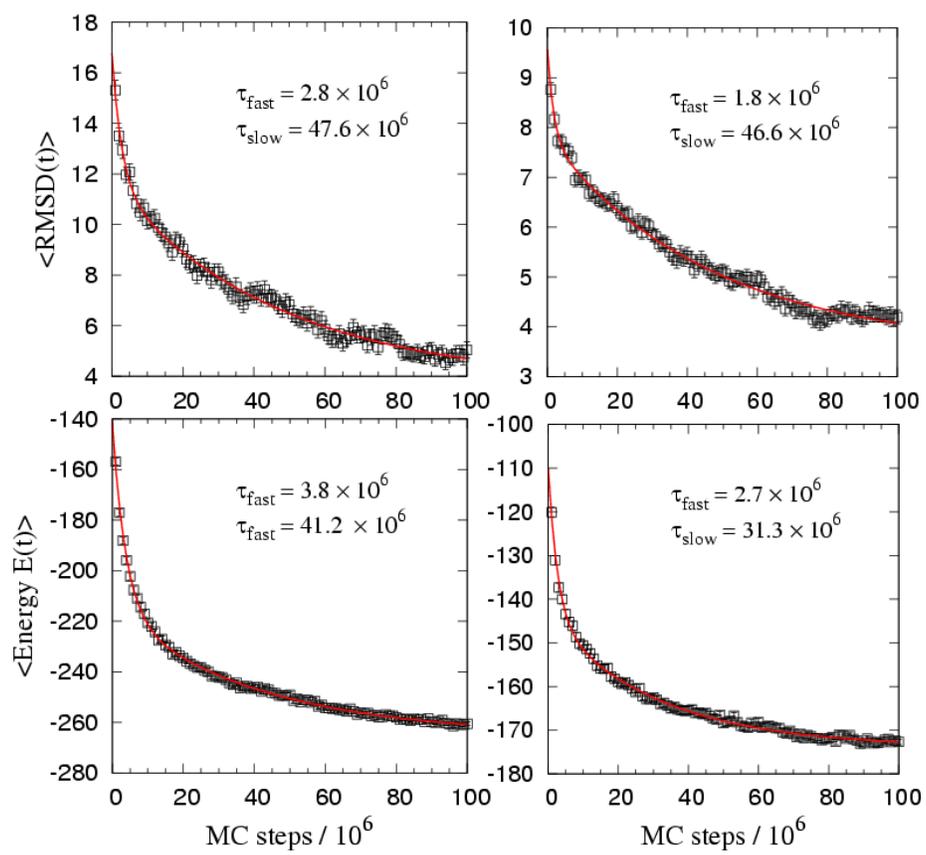

**Figure S3**

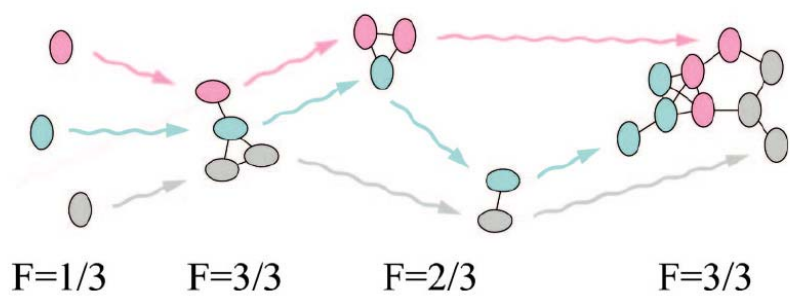

F=1/3          F=3/3          F=2/3          F=3/3

**Figure S4**

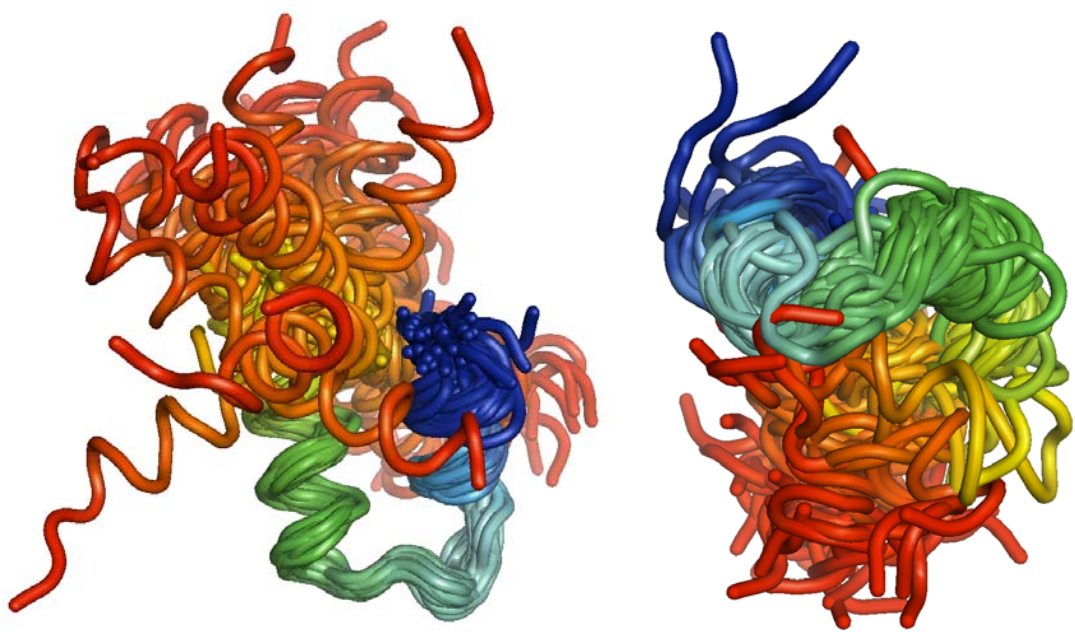

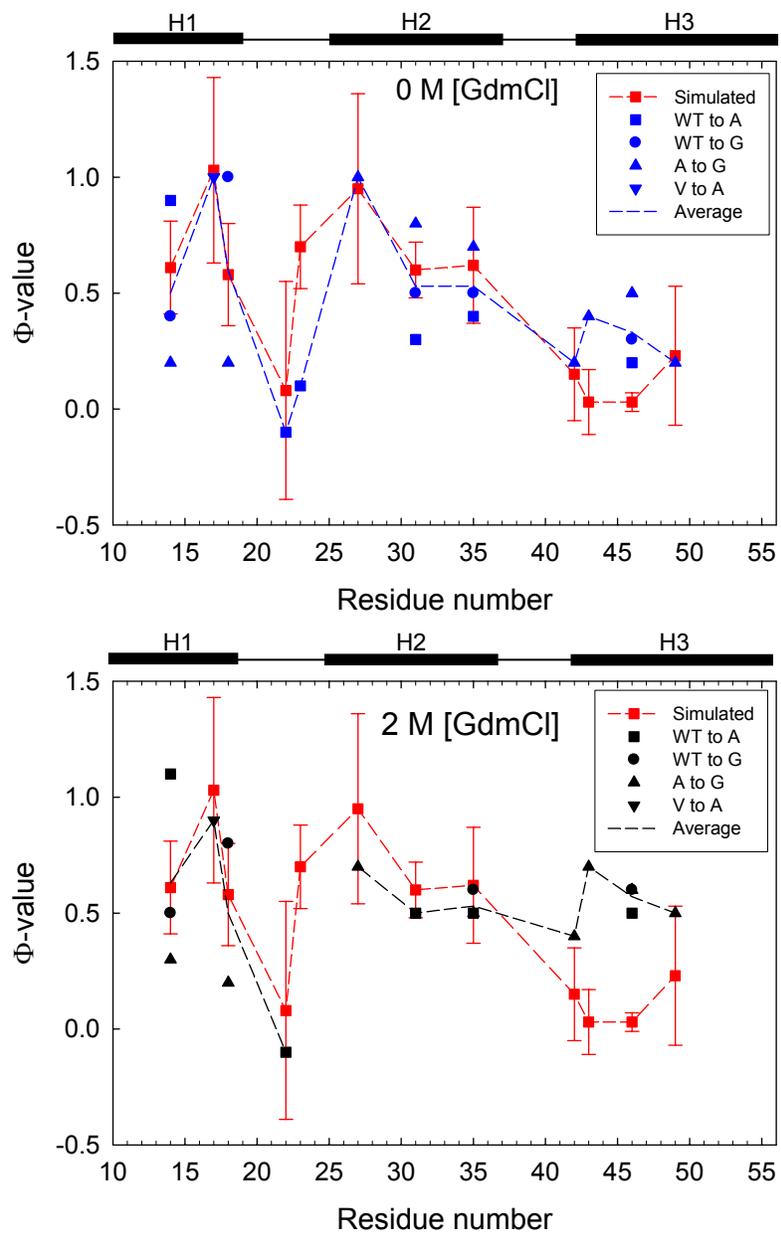

**Figure S6**

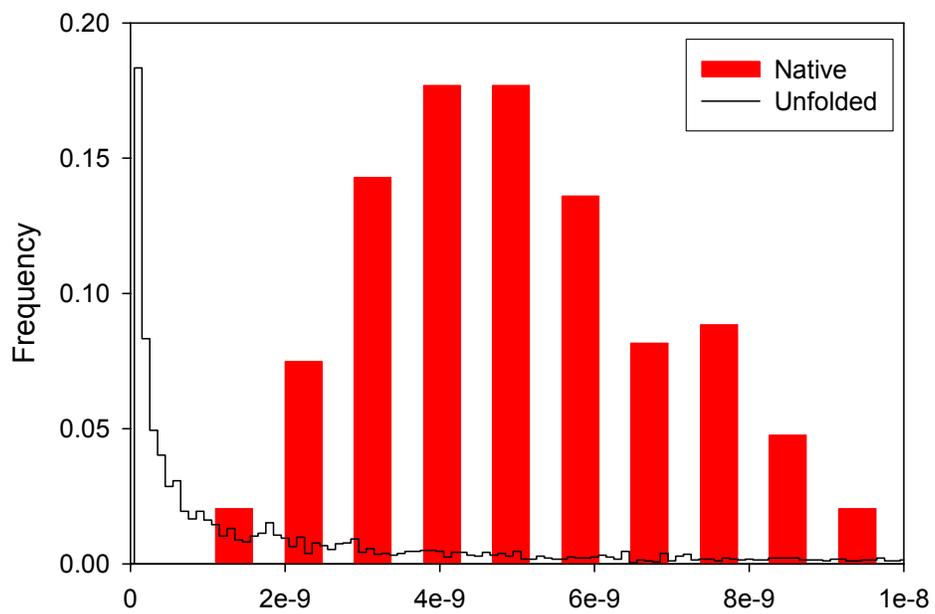

$$1/r^6 \; (\text{Å}^{-6})$$

**Figure S7**

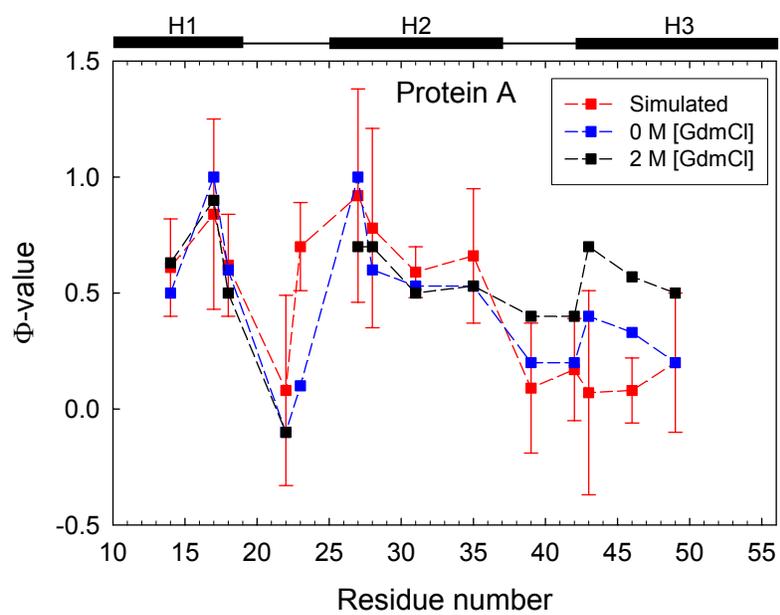

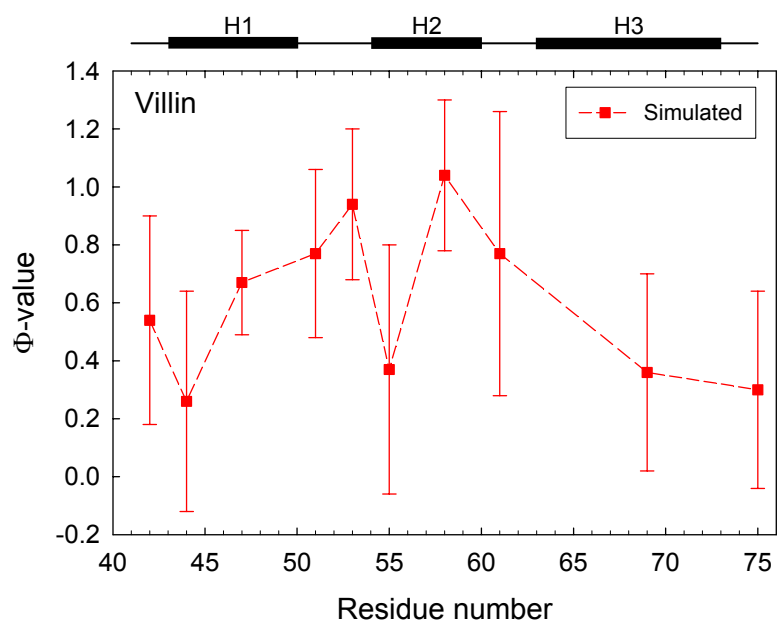